# The New Frontier of Cybersecurity: Emerging Threats and Innovations

[1] Daksh Dave, [2] Gauransh Sawhney, [3] Pushkar Aggarwal, [4] Nitish Silswal, [5] Dhruv Khut

1 Department of Electrical and Electronics, BITS Pilani, Pilani, India
2 Department of Computer Science and Engineering, BITS Pilani Goa Campus, Goa, India
3 Department of Information Technology, Sardar Patel Institute of Technology, Mumbai, India

*Abstract.* In today's digitally interconnected world, cybersecurity threats have reached unprecedented levels, presenting a pressing concern for individuals, organizations, and governments. This study employs a qualitative research approach to comprehensively examine the diverse threats of cybersecurity and their impacts across various sectors. Four primary categories of threats are identified and analyzed, encompassing malware attacks, social engineering attacks, network vulnerabilities, and data breaches. The research delves into the consequences of these threats on individuals, organizations, and society at large. The findings reveal a range of key emerging threats in cybersecurity, including advanced persistent threats, ransomware attacks, Internet of Things (IoT) vulnerabilities, and social engineering exploits. Consequently, it is evident that emerging cybersecurity threats pose substantial risks to both organizations and individuals. The sophistication and diversity of these emerging threats necessitate a multi-layered approach to cybersecurity. This approach should include robust security measures, comprehensive employee training, and regular security audits. The implications of these emerging threats are extensive, with potential consequences such as financial loss, reputational damage, and compromised personal information. This study emphasizes the importance of implementing effective measures to mitigate these threats. It highlights the significance of using strong passwords, encryption methods, and regularly updating software to bolster cyber defenses.

## I. INTRODUCTION

In our rapidly evolving and increasingly digitalized world, cybersecurity has emerged as a paramount concern, affecting individuals, organizations, and governments alike [1]. As technology advances at an unprecedented pace, so do the cyber threats and attacks that target our interconnected systems. To effectively safeguard against these malicious actors, it is crucial to continually monitor and analyze the emerging cybersecurity threats that pose significant risks. The field of cybersecurity, encompassing an array of measures and technologies, aims to protect computers, servers, mobile devices, electronic systems, networks, and data from digital attacks and unauthorized access [2]. However, the ever-expanding capabilities of cybercriminals and the relentless discovery of new threats and vulnerabilities present an ongoing challenge [3]. This research embarks on a comprehensive exploration of the multifaceted world of cybersecurity threats, with a sharp focus on the emerging ones that carry substantial implications for individuals and organizations alike.

The study will delve into various menacing types of threats, ranging from well-known malware and phishing attacks to increasingly sophisticated ransomware and social engineering exploits. Moreover, to gain a deeper understanding of the threats' underlying motivations, the investigation will delve into the tactics employed by attackers to exploit weaknesses in systems and networks. By analyzing recent trends in cyber-attacks, the study aims to assess the potential consequences on individual privacy, financial security, and overall organizational stability. Effectively countering and mitigating these emerging threats demands a comprehensive understanding of their dynamics [4]. Organizations can proactively implement robust security measures and impart employee training to foster a cybersecurity-aware culture that safeguards sensitive information [5]. Simultaneously, individuals can take proactive steps to enhance their personal cybersecurity, such as employing strong passwords, keeping software up to date, and exercising caution when encountering suspicious emails or websites [6].

A successful defense against these rapidly evolving threats requires cybersecurity professionals to stay abreast of the latest trends, vulnerabilities, and attack techniques [7]. Emphasizing a multi-layered defense approach becomes imperative, combining advanced technological solutions, informed employee training, and proactive threat intelligence. Additionally, effective collaboration among governments, businesses, and individuals plays a pivotal role in developing robust cybersecurity strategies and sharing crucial information on emerging threats [8]. Through this study, we aim to shed light on the ever-changing landscape of cybersecurity threats and provide invaluable insights into preventive and mitigative measures. By remaining informed and vigilant, we can collectively work towards forging a safer and more secure digital future.

## II. LITERATURE REVIEW

### A. Taxonomy of Approaches Used For Extracting The Relevant Literature Review

To carry out this literature review, a comprehensive search was conducted on various online databases, including IEEE Xplore, ACM Digital Library, and Google Scholar. Keywords such as "emerging threats," "cybersecurity," "cyber attacks," and "cyber threats" were used to identify relevant research papers and articles. The search was limited to peer-reviewed publications published within the last five years.

## B. Cybersecurity and Its Importance

Cybersecurity is a critical activity aimed at preventing unauthorized access, disclosure, interruption, modification, and destruction of computer systems, networks, and data [9]. It involves implementing robust procedures to identify and thwart various online dangers, such as malware, phishing scams, ransomware, and other cybercrimes. To achieve this, cybersecurity encompasses safeguarding hardware, networks, and data, along with educating users, enforcing security policies, and maintaining incident response plans for cyberattack scenarios [10]. The goal of cybersecurity is to ensure the availability, confidentiality, and integrity of data and systems [11].

Cybersecurity's importance in our interconnected world has grown exponentially with the proliferation of technology and the widespread use of the internet, leading to severe consequences like financial loss, reputational damage, and even loss of life in some cases due to cyber-attacks [12]. Protecting personal information is a key reason why cybersecurity is crucial in today's digital age, where almost everyone has an online presence through various activities such as social media, online banking, and shopping, making their data constantly at risk of being targeted by cybercriminals [13]. Implementing robust cybersecurity measures is essential for enterprises and organizations, as cyberattacks can lead to significant financial losses and reputational harm, particularly devastating for small and medium-sized organizations without adequate resources for recovery [14]. Moreover, cybersecurity is crucial for fostering trust and confidence in the digital economy, where consumers must feel assured that their information remains secure, encouraging online engagement and supporting economic growth and innovation [15].

## C. Evolution of Cybersecurity Threats

Cybersecurity threats have been a concern since the early days of computer networks. As technology advanced and the Internet became more widespread, the number and complexity of these threats increased dramatically [4]. Here is an overview of the history of cybersecurity threats:

*Early Years (1970s-1990s):* The first cybersecurity threats emerged in the 1970s with the advent of early computer networks. Hackers began exploiting vulnerabilities in operating systems and software, often for the thrill or to gain unauthorized access. These early threats were mostly isolated incidents, and security measures were minimal [3].

*Viruses and Malware (1990s-2000s):* The 1990s saw the rise of computer viruses and malware. These were often spread through infected floppy disks or email attachments. The infamous Michelangelo virus, released in 1991, infected thousands of computers worldwide. Cybercriminals also started using worms and trojans to gain unauthorized access and steal information [16].

*Growing Internet and Web-Based Threats (2000s-2010s):* With the growth of the Internet and the widespread adoption of web-based technologies, new types of threats emerged. Phishing attacks, where attackers tricked users into revealing sensitive information through deceptive emails, became prevalent [17]. Distributed Denial of Service (DDoS) attacks, where multiple computers bombard a target website or server with traffic, leading to an overload and temporary shutdown, became a significant threat as well [17].

*Advanced Persistent Threats (2010s-present):* In recent years, advanced persistent threats (APTs) have become a major concern. APTs are sophisticated and targeted attacks, often sponsored by nation-states, aimed at compromising specific individuals, organizations, or industries [19]. These attacks involve a combination of social engineering, malware, and network exploitation techniques. APTs can be difficult to detect and mitigate due to their stealthy nature [15].

*Ransomware and Extortion (2010s-present):* Ransomware attacks have seen a significant increase in recent years. Ransomware is a type of malware that encrypts a victim's data, rendering it inaccessible until a ransom is paid to the attacker [20]. Cybercriminals have targeted organizations of all sizes, including healthcare providers, governments, and educational institutions. These attacks can have severe consequences, including financial loss and disruption of critical services [9].

*Internet of Things (IoT) Threats:* As more devices become connected to the Internet, there is growing concern about the security of the Internet of Things (IoT). IoT devices, including smart appliances, wearable technology, and home automation systems, can be vulnerable to cyberattacks [21]. These attacks can range from gaining unauthorized access to compromising personal data or even controlling physical devices. The inherent weaknesses in IoT device security and the large-scale deployment of these devices make them an attractive target for cybercriminals [11].

*Cloud-Based Threats:* With the widespread adoption of cloud computing, new cybersecurity threats have emerged. Cloud-based attacks can involve compromising cloud storage accounts, exploiting vulnerabilities in cloud infrastructure, or stealing sensitive data from cloud-based applications [22]. As more organizations move their data and operations to the cloud, ensuring the security of cloud-based systems and data has become crucial [23].

## III. SOURCES OF CYBER THREATS

Cyber threats can originate from a diverse array of sources, encompassing individuals, criminal organizations, nation-states, and even insiders within an organization.

- Individuals, often known as "script kiddies," leverage free or inexpensive internet tools to launch attacks driven by motives such as curiosity, seeking recognition, or causing disruption [24].
- Criminal organizations constitute another significant source of cyber threats, motivated by financial gain and involved in various activities, including credit card theft, ransomware attacks, and identity theft [25].
- Nation-states possess advanced cyber warfare capabilities, making them formidable sources of cyber threats. They engage in a wide spectrum of activities, ranging from intelligence gathering to the potential disruption or destruction of critical infrastructure [26].
- Insiders within an organization can also pose considerable cyber threats, exploiting their access to sensitive information or systems and using their insider knowledge to launch attacks or compromise valuable assets [27].

TABLE 1. TRENDS IN CHANGING CYBERCRIME ATTACKS (IN MILLIONS).

| Incidents | 2015 | 2016 | 2017 | 2018 | 2019 | 2020 | 2021 | 2022 | %Increase /Source |
|---|---|---|---|---|---|---|---|---|---|
| Fraud Intrusion | 3.430 | 3.546 | 3.640 | 3.767 | 3.830 | 3.945 | 4.001 | 4.040 | 18% [30] |
| Fraud | 3.439 | 3.545 | 3.737 | 3.879 | 4.157 | 4.456 | 4.746 | 5.191 | 50% [10] |
| Spam | 0.320 | 0.410 | 0.617 | 0.714 | 0.870 | 1.456 | 1.765 | 1.956 | 511% [32] |
| Denial ofService | 0.367 | 0.544 | 0.676 | 0.754 | 0.945 | 1.276 | 1.382 | 1.568 | 327% [10] |
| Cyber Harassment | 0.345 | 0.454 | 0.655 | 0.764 | 0.845 | 1.156 | 1.266 | 1.478 | 328% [32] |
| Vulnerabilityreports | 0.310 | 0.515 | 0.614 | 0.811 | 0.915 | 0.990 | 1.372 | 1.491 | 381% [37] |
| IntrusionAttempts | 0.257 | 0.417 | 0.614 | 0.715 | 0.791 | 1.154 | 1.256 | 1.382 | 438% [28] |
| Maliciouscode | 0.410 | 0.545 | 0.756 | 0.967 | 1.176 | 1.245 | 1.444 | 1.580 | 285% [37] |
| Contentrelated | 0.201 | 0.375 | 0.576 | 0.847 | 0.944 | 1.176 | 1.276 | 1.364 | 578% [28] |
| **Total** | **9.079** | **10.351** | **11.298** | **13.218** | **14.473** | **16.854** | **18.508** | **20.05** | |

Table 1 above shows the number of incidents reported for various types of cyber threats from 2015 to 2022 and the percentage increase each year.

The data in the table shows an overall increase in cybercrime attacks over the years. For most of the categories, there is a gradual increase in the number of incidents reported each year. Analyzing the trends in cybercrime attacks, we can see that most types of cybercrime attacks have experienced an increase in incidents over the years. Fraud intrusion, fraud, denial of services, cyber harassment, vulnerability reports, intrusion attempts, malicious code, and content-related incidents have all shown an upward trend in the number of incidents. For example, fraud intrusion incidents increased from 3.430 million in 2015 to 4.040 million in 2022, showing an 18% increase. Similarly, fraud incidents increased from 3.439 million in 2015 to 5.191 million in 2022, showing a 50% increase. Overall, the total number of cybercrime attacks has steadily increased from 9.079 million incidents in 2015 to 20.05 million incidents in 2022. This represents a significant 121% increase in cybercrime attacks over the seven-year period. It is important to note that these numbers represent reported incidents and may not capture the full extent of cyber threats as many incidents go unreported.

The increase in cybercrime attacks can be attributed to advancements in technology, increasing connectivity, economic incentives, lack of cybersecurity measures, and evolving tactics and techniques used by cybercriminals. It is crucial for individuals, organizations, and governments to invest in robust cybersecurity measures to mitigate the risks and combat the growing threat of cybercrime.

## IV. EMERGING THREATS IN CYBERSECURITY

### A. Malware

Malware, short for malicious software, encompasses any software designed to harm or exploit computers and networks, including viruses, worms, Trojans, ransomware, spyware, adware, and other malicious programs [12]. This insidious software is typically disseminated through email attachments, infected websites, software downloads, or exploits in vulnerable software [28]. Once infiltrating a computer, malware can wreak havoc by stealing sensitive information, corrupting or deleting files, disrupting system operations, and granting unauthorized access to networks. To safeguard against these threats, it is crucial to have robust antivirus and anti-malware software installed on your computer [29].

TABLE 2. MALWARE DEFINITION

| Malware | Categories | Mitigation practices | Infection | How it spreads |
|---|---|---|---|---|
| Computer worm | Morris worm Blaster worm Black worm | Antivirus Software update OS and software update | Irregularities in the web browser System and OS error fault | Sharing of files Emails |
| Computer virus | Retro virus Macro virus Boot sector virus | Use antivirus software Conducting scanning periodically | Destruction of system hardware and program file Reduces loading speed in the host computer | Email attachment Downloads from internet |
| Trojan horse | Zlob Trojan Coreflood | Anti-Trojan program Use antivirus software | Theft of credit cards and passwords System breakdown | Movies Image games Mp3 files |

### B. Social Engineering

Social engineering is a manipulative tactic employed by attackers to deceive individuals into revealing private information or engaging in actions that compromise security [29]. By exploiting human behavior and gaining trust, attackers aim to extract personal data like passwords or financial information or manipulate victims into performing actions they would not typically do, such as opening malicious email attachments or clicking on harmful links [30].This form of attack manifests in various guises, including phishing emails or phone calls impersonating legitimate organizations or individuals, as well as forging relationships by posing as coworkers or IT personnel. Attackers may also create fraudulent websites or social media profiles to establish trust [31]. In-person tactics like tailgating (unauthorized entry into secure areas behind authorized personnel) and pretexting (using fabricated scenarios to gain trust and extract information) are also part of the social engineering arsenal [18].

The objectives of social engineering attacks can range from identity theft and financial fraud to gaining unauthorized access to systems or networks and even disseminating malware or ransomware [17]. Safeguarding against these attacks requires heightened awareness, skepticism, and education. It is crucial to exercise caution when sharing personal information or engaging in actions involving security risks. Recognizing common social engineering techniques and identifying warning signs empowers individuals to avoid falling prey to such schemes. For organizations, implementing robust security measures involves educating employees about recognizing and responding to social engineering attacks, employing multi-factor authentication, and regularly updating security protocols and systems [2]. By fostering a security-conscious culture and bolstering defense strategies, individuals and organizations can fortify their resilience against social engineering threats.

### C. Phishing

Phishing is a prevalent method of cybercrime wherein attackers masquerade as reputable entities, such as banks or online service providers, with the intent to deceive individuals into divulging sensitive information, like passwords or credit card numbers [7]. This deceit is commonly achieved through fraudulent emails or counterfeit websites that closely resemble legitimate ones. Once the attackers acquire the stolen information, they can exploit it for malicious purposes, including identity theft or financial fraud. Phishing attacks pose a significant threat to both individuals and organizations, underscoring the importance of recognizing phishing indicators, such as suspicious email requests or unfamiliar website URLs, to safeguard against falling victim to these scams [15].

### D. Ransomware

Ransomware, a malicious form of software, employs encryption to render a victim's data inaccessible and subsequently demands payment in exchange for the decryption key. This malevolent tactic targets individuals, businesses, and even governments alike. Ransomware attacks often ensue through phishing emails, malicious downloads, or exploiting vulnerabilities in software or systems [3]. Once the victim's data is encrypted, the ransomware displays a message instructing them on how to pay the ransom, typically in the form of cryptocurrency like Bitcoin. In the event of non-payment within a specified timeframe, the attackers may resort to threats of erasing the decryption key or permanently deleting the encrypted contents [6].

The consequences of ransomware attacks can be severe, leading to financial loss, data breaches, and even the disruption of critical infrastructure. To mitigate such risks, it is crucial to maintain regular backups of files and systems, keep software up-to-date, and exercise caution when encountering suspicious emails or downloads [9]. By adopting these proactive measures, individuals and organizations can enhance their defenses against ransomware attacks and protect their valuable data and assets.

### E. Denial of Service (DoS) Attacks

A denial of service (DoS) attack is a type of cyberattack wherein an individual deliberately inundates a network, server, or website with excessive traffic or data to exhaust its resources, rendering it inaccessible to users [4]. The objective of a DoS attack is to disrupt the normal functioning of the target system or network, causing unavailability or crashes [11].
DoS attacks can be executed through various methods, such as overwhelming the target with a high volume of network requests, exploiting vulnerabilities in the target's software or infrastructure, or utilizing botnets - networks of compromised computers - to inundate the target with traffic from multiple sources [32].

To effectively combat DoS attacks, organizations should have a well-defined incident response plan in place. This plan should facilitate swift actions to mitigate the attack's effects and minimize potential damage [33]. Key components may include procedures for isolating affected systems, promptly notifying relevant stakeholders, and collaborating with law enforcement agencies if necessary [10].

### F. Zero-Day Exploits

Zero-day exploits refer to undisclosed software or hardware vulnerabilities that are unknown to developers or manufacturers [8]. Malicious actors can exploit these vulnerabilities to gain unauthorized access to systems or networks, giving them a significant advantage as developers have not had a chance to patch or fix them, hence the term "zero-day." Such exploits are highly coveted due to their potential to penetrate systems undetected. The discovery and use of zero-day exploits can result in severe consequences, necessitating organizations to remain vigilant in their efforts to defend against these threats [31].

### G. Advanced Persistent Threats (APTs)

Advanced Persistent Threats (APTs) represent targeted cyber-attacks executed by well-funded and highly skilled hackers, often nation-state actors, aiming to gain unauthorized access to a specific organization's network or system [34]. APTs are characterized by their persistence, as attackers may prolong their operations to achieve their objectives over an extended period. The stages of APTs typically encompass initial reconnaissance and infiltration, establishing footholds, lateral movement through the network, and exfiltration or manipulation of sensitive data [33]. To evade detection, attackers employ sophisticated techniques like custom malware, zero-day vulnerabilities, and social engineering tactics [35]. Motives driving APTs can range from intelligence gathering and stealing intellectual property to conducting espionage or disrupting critical infrastructure. The targeted organizations are typically entities with valuable and sensitive information, including government agencies, defense contractors, financial institutions, and research and development companies [24] .

Effectively protecting against APTs requires implementing comprehensive security measures, such as network segmentation, strong access controls, regular vulnerability assessments, and educating employees on security best practices. Utilizing threat intelligence and advanced detection technologies can aid in identifying and responding to APTs promptly. A proactive and multi-layered defense strategy is crucial to safeguarding organizations from the persistent and advanced nature of APTs [36].

## V. STRATEGIES TO MITIGATE CYBERSECURITY THREATS

### A. Regular Software Updates

One of the most effective strategies to mitigate cybersecurity threats is regular software and security patch updates. Software developers release updates to address vulnerabilities and bugs that hackers may exploit, ensuring users have the latest security measures in place [3]. Cybercriminals constantly evolve tactics and techniques, seeking vulnerabilities in popular software applications, operating systems, and plugins to exploit. Failing to update software leaves users vulnerable to evolving threats, while consistent updates help stay ahead of cybercriminals and close security loopholes [37].

Regular software upgrades not only fix bugs but also address security vulnerabilities and improve system efficiency, preventing system failures and data loss caused by outdated or incompatible software [9]. Updating software is easy through automatic updates or manual checks in software settings. Enabling automatic updates is recommended to ensure the software is always up to date without requiring active user involvement [35]. Besides software updates, regularly updating the operating system, web browsers, antivirus software, and other applications is crucial. Neglecting these updates exposes users to known security flaws that developers have already addressed [12].

### B. Implementation of Firewalls and Intrusion Detection Systems

A firewall is a network security device that tracks incoming and outgoing network traffic, allowing or blocking certain traffic based on predefined security rules. Employing firewalls enables businesses to separate their internal network from the public internet, safeguarding critical information [29]. In addition to firewalls, organizations should implement intrusion detection systems (IDS), which monitor network traffic for suspicious and malicious activity, detecting anomalies and potential security breaches. IDS alerts system administrators or security personnel when an intrusion is detected, enabling immediate action [1].

Firewalls and IDS work together to identify and block potential threats, providing an additional layer of protection against cyber-attacks. By controlling network access and responding to suspicious activity, organizations can significantly reduce the risk of cybersecurity threats [14]. However, implementing firewalls and IDS alone is insufficient for comprehensive cybersecurity. Organizations should also adopt a multi-layered security approach, including strong passwords, encryption for sensitive data, software patching, employee training, and data backups [37]. By integrating firewalls and IDS and embracing a multi-layered security approach, organizations can significantly decrease their vulnerability to cybersecurity threats.

### C. Incident Response Planning

An incident response plan outlines the actions an organization will take in the event of a cybersecurity issue. This includes locating, containing, and investigating the incident to determine its scope. The plan also guides appropriate actions, such as alerting relevant parties, data recovery, and implementing safeguards to prevent future incidents [13]. A well-developed incident response plan enables organizations to address cybersecurity incidents promptly and efficiently, minimizing the breach's impact and preventing further damage. Roles, responsibilities, and communication channels should be clearly outlined, along with guidance for responding to different incident types. Regular testing and updates are essential to maintain plan effectiveness [32].

### CONCLUSION

This comprehensive review article offers a detailed overview of the emerging cybersecurity threats, encompassing various types of risks such as malware, ransomware, phishing attacks, and social engineering. It delves deep into the motivations behind cyber-attacks, ranging from seeking financial gain to pursuing political and espionage objectives. Moreover, the study underlines the broad-reaching impact of these threats on individuals, businesses, and governments, highlighting the urgent need for robust cybersecurity measures. The potential consequences of cyber-attacks are far-reaching, spanning from significant financial losses and reputational damage to the compromise of national security. In response to these dynamic threats, the review emphasizes the importance of maintaining constant vigilance and adopting proactive measures. Cybersecurity professionals must continually update their knowledge and skills to stay ahead of the ever-evolving tactics employed by cybercriminals. Additionally, the study shines a light on the role of technology, which acts as a double-edged sword. While advancements in technology introduce new vulnerabilities, they also offer opportunities for innovation in cybersecurity solutions.